\documentclass[acmsmall]{acmart}

\AtBeginDocument{%
  \providecommand\BibTeX{{%
    \normalfont B\kern-0.5em{\scshape i\kern-0.25em b}\kern-0.8em\TeX}}}

\usepackage{booktabs} 

\usepackage{bm}
\usepackage{tikz}
\usetikzlibrary{external,automata,trees,positioning,shadows,matrix,arrows,shapes.geometric,shapes.multipart,trees,calc,fit,decorations.pathmorphing}
\usetikzlibrary{positioning,shapes,decorations,backgrounds}
\usetikzlibrary{matrix}
\usepackage{pgfplots}
\pgfplotsset{compat=1.14}
\usepackage{bchart}

\pgfdeclarelayer{background}
\pgfdeclarelayer{foreground}
\pgfsetlayers{background,main,foreground}	

\usetikzlibrary{calc,trees,positioning,arrows}
\usetikzlibrary{automata,positioning}
\usetikzlibrary{trees}
\usetikzlibrary{decorations.pathmorphing}	
\usetikzlibrary{fit}	
\usetikzlibrary{backgrounds} 	
\usetikzlibrary{positioning}
\usetikzlibrary{shadows}
\usepgflibrary{shapes}

\tikzstyle{small}=[circle, thick, minimum size=.5cm, draw=black!80]
\tikzstyle{square}=[style={regular polygon,regular polygon sides=4}, inner sep=-1.3em, thick, minimum size=.25cm, draw=black!100, fill=black!100]
\tikzstyle{plate}=[rectangle, thick, inner sep=0.25cm, draw=black!100]
\tikzstyle{shadeplate}=[rectangle, thick, inner sep=0.4cm, draw=black!100]
\tikzstyle{table}=[circle,fill=blue!20,draw=black!100,inner sep=1pt, minimum size=30pt]
\tikzstyle{client}=[rectangle,fill=blue!20,draw=black!100,inner sep=1pt, minimum size=12pt]
\usepackage{multirow}

\usepackage[ruled]{algorithm2e} 

\SetAlFnt{\small}
\SetAlCapFnt{\small}
\SetAlCapNameFnt{\small}
\SetAlCapHSkip{0pt}
\IncMargin{-\parindent}

\usepackage{centernot}
\newcommand\independent{\protect\mathpalette{\protect\independenT}{\perp}}
\def\independenT#1#2{\mathrel{\rlap{$#1#2$}\mkern2mu{#1#2}}}

\RequirePackage[normalem]{ulem} 
\RequirePackage{color}\definecolor{RED}{rgb}{1,0,0}\definecolor{BLUE}{rgb}{0,0,1} 

\begin{document}
\title{Can Smartphone Co-locations Detect Friendship? It Depends How You Model It.}

 \author{Momin M. Malik}
 \authornote{This is the corresponding author}
 \orcid{0000-0002-4871-0429}
 \affiliation{%
   \institution{Harvard University}
   \state{MA}
   \country{USA}}
 \email{momin.malik@gmail.com}

 \author{Afsaneh Doryab}
 \affiliation{%
   \institution{University of Virginia}
   \department{School of Engineering and Applied Science}
  \state{VA}
  \country{USA}
 }
 \email{ad4ks@virginia.edu}

 \author{Michael Merrill}
 \affiliation{%
   \institution{University of Washington}
  \state{WA}
 }
 \email{mam546@cornell.edu}

 \author{J\"{u}rgen Pfeffer}
 \affiliation{%
  \institution{Technical University of Munich}
   \department{Bavarian School of Public Policy}
   \country{Germany}
 }
 \email{juergen.pfeffer@tum.de}

 \author{Anind K. Dey}
 \affiliation{%
  \institution{University of Washington}
   \department{Information School}
   \state{WA}
   \country{USA}
 }
 \email{anind@uw.edu}

\begin{abstract}
We present a study to detect friendship, its strength, and its change from smartphone location data collected among members of a fraternity. We extract a rich set of co-location features and  build classifiers that detect friendships and close friendship at 30\% above a random baseline. We design cross-validation schema to test our model performance in specific application settings, finding it robust to seeing new dyads and to temporal variance. 
\end{abstract}

%
%


\maketitle

\section{Introduction}
Friendship is a ``voluntary, personal relationship typically providing intimacy and assistance'', associated with characteristics of trust, loyalty, and self-disclosure \cite{fehr1996}. It is one of the most important aspects of human existence, lending meaning to life, and providing for material, cognitive, and social-emotional needs in ways that lead to greater health and well-being \cite{fehr1996}. 

Understanding the friendship relationship between people can be helpful for creating technology that serves people better. If an individual's friendships are known, these can be leveraged for applications supporting help-seeking behavior such as requests for recommendations or for favors, or for automatically establishing trust between users' devices. Friendships may also be leveraged for carrying out social interventions around diet and exercise \cite{madan2010} or for preventing disease transmission \cite{madan2010b}, such as in mobile applications that facilitate friends adopting and holding each other accountable to healthy behaviors. Conversely, it may be of interest to create applications to make recommendations about friendships in order to help bring people together, for example at conferences \cite{chin2012,chin2013} or for information dissemination in workplaces \cite{lawrence2006}. Longitudinal information about changes in friendships could help detect the onset of isolation and help design interventions to strengthen friendships. 

But in order to incorporate information about one's friendship network in personal informatics and mobile applications, we need ways of detecting friendship.\footnote{Detecting friendship is a link prediction problem \cite{liben2007}, although we use the term `detection' to emphasize that our predictions are of concurrent, not future, values. We avoid `infer' as we do not employ standard errors.} One easy way to get `ground truth' is to rely on ties from online social networking platforms; but such ties are not necessarily good proxies for the underlying construct of friendship \cite{wilson2012,corten2012,mislove2007}. Survey instruments have been the standard network data collection method in social network analysis for decades, but involve a high burden for users that make them impractical as a basis for mobile applications. 

Previous social science work has established the strong link between individuals being physically close and being friends \cite{festinger1950,fehr1996,latane1995}. There is a two-way causal connection in that people who spend time together are more likely to become friends \cite{festinger1950}, and that people who are friends spend time together \cite{fehr1996}. This suggests that there should be a robust signal of friendship in measurements of physical proximity that can then be leveraged for analysis, services, or interventions. 

Our work is the first effort to detect friendships using feature extraction from smartphone location data. Previous works either descriptively, rather than predictively, linked location data and friendship \cite{eagle2009}, looked at ties on location-based online social network services \cite{cranshaw2010}, or used mobile phone call and SMS logs \cite{wiese2014,wiese2015}. We believe that detecting friendship from mobile phone co-location data is a realistic approach for future mobile applications and interventions that seek to leverage friendship for other tasks. 

This paper presents the results of a 3-month study of a cohort of 53 participants, with final analysis performed on 9 weeks of data from 48 participants. We combine mobile phone sensor data collection with established social network survey instruments, and use rich feature extraction from co-location data to see how well such data can be used to detect friendships, {\it close} friendships, and {\it changes} in friendship. 

Our contributions are as follows:
\begin{itemize}
\item We present, to our knowledge, the first {\it pairwise} feature extraction from smartphone location data, and show that a classifier built with the extracted features performs 30\% above random (Matthews correlation coefficient). This can serve as a baseline for all future work. 
\item We design a novel evaluation method (using temporal block assignment cross-validation and what we call dyadic assignment cross-validation ) to mimic different realistic application settings in order to more rigorously test our classification's generalizability to these settings, and use it to show that our approach is robust to seeing new pairings of individuals, and to variability in co-location patterns over time. 


\end{itemize}

Below, first we review background work in social network analysis and in the mobile and pervasive computing literature around existing approaches to extracting interactions and/or friendships from mobile sensor data. We describe our study design, and our feature extraction process. With extracted features, we build a predictive model and use feature importance as a step towards characterizing the aspects of co-location most useful for detecting friendship. 

\section{Background and related work}
\subsection{Friendship ties}
Friendship is an example of a \emph{relational phenomenon}. Relational phenomena are two-unit or \emph{dyadic} relations
\cite{borgatti2009}. Instead of the $n$ individuals of a dataset being the observations, they have as observations $\binom{n}{2}$ undirected (symmetric) relations (e.g., co-location), or $2 \times \binom{n}{2}$ directed (potentially asymmetric) relations (e.g.,  self-reported friendships) between individuals. 

Research throughout the 20th century has provided examples of friendship and other ties having explanatory and predictive power, such as in explaining why girls ran away from a school for delinquent teenage girls and predicting future runaways \cite{moreno1934}, explaining the breakup of a monastery \cite{sampson1968}, of the split of a karate club into two separate clubs \cite{zachary1977}, or using structures of informal networks to predict the success or failure of institutional reorganization \cite{krackhardt1996}. Insights from these approaches have proved robust as they have been successfully integrated into search engines, recommender systems, and the structure of social media platforms. 

We represent a collection of friendship ties between $n$ people as an $n \times n$ adjacency matrix, $A$, where 
\[
A_{ij} = \begin{cases}
1 \mbox{ if there is a tie } i \to j \\
0 \mbox{ otherwise.}
\end{cases}
\]
and where $A_{ii} = 0$ (no self-loops). $A$ is not necessarily symmetric, as friendship ties collected from sources like surveys can and do yield cases of $A_{ij} \neq A_{ji}$. As an (asymmetric) adjacency matrix defines a (directed) graph, which in this case is a {\it friendship network}, we refer to the $n$ people as nodes. We refer to an (unordered) pair of individuals $(i, j)$ as a {\it dyad}, and a value where  $A_{ij} = 1$ as an {\it edge}, and also interchangeably as a {\it tie} or {\it link}.  

While $A_{ij}$ represents a `dependency' between units $i$ and $j$, for example $A_{ij}=1$ will frequently be associated with similar outcomes from $i$ and $j$, such ties are themselves dependent \cite{snijders2011}. For example, while friendship ties are  not necessarily mutual or {\it reciprocated}, we still have $A_{ij} = A_{ji}$ more often than we would expect at random. In symbolic terms, $P(A_{ij}) \centernot{\independent} P(A_{ji})$

Other such dependencies include {\it transitivity}, which are friend-of-a-friend connections, $P(A_{ij}) \centernot{\independent} P(\{A_{ik}, A_{kj} : k \neq i, j\})$, and preferential attachment \citep{perc2014}, $P(A_{ij} \centernot{\independent} P(\{A_{kj} : k \neq j\})$. 

Not accounting for such dependencies can, in explanatory models, lead to too-small standard errors and omitted variable bias \cite{dow1982b,dow1984,dow2007}. In predictive  models \cite{shmueli2010,breiman2001}, dyadic dependencies impact the validity of cross-validation estimates of model performance \cite{dabbs2016} in ways similar the impacts of other types of dependencies \cite{hammerla2015}.

For example, if $A_{ij} = A_{ji}$, for co-location features $X_{ij}$, if the pair $(A_{ij}, X_{ij})$ is in the training set and $(A_{ji}, X_{ij})$ is in the test set, we would be training and testing on the exact same row of data! 

In any link prediction task, having a cross-validation scheme that does not reflect an application setting (for example, using features related to network degree or to the number of mutual friends when we may not have the complete network) can potentially give a misleading picture of what true out-of-sample performance would be. 

To avoid these problems, we first avoid using features that would not be available in an application settings (network features, lagged values of the class label, etc.). We also employ three different cross-validation schemes, described below. There are existing methods for doing cross validation with network data \cite{chen2017,dabbs2016}, but their validity depends on strong assumptions and even then they do not control for all known dependencies; instead, by employing different cross validation schema, we get a better picture of how our classifier would work in different use cases.

\subsection{Friendship and co-location}
The connection of friendship and proximity has long been a topic of study in social science. In a foundational work of social psychology carried out in 1946, the `Westgate study', Festinger et al. \cite{festinger1950} carried out a field experiment around soldiers returning from WWII and attending graduate school at MIT on the GI bill. They and their families were randomly placed into the units of the relatively isolated, newly built Westgate housing complex. The study authors were able to quantify the extent to which people living close by, or passing one another on the way to their residences, were more likely to become friends than with others with whom they did not have opportunities for interaction (although the study looking only at men may have neglected an important causal process in the role of women, \cite{cherry1995}). Then, in 1954 and 1955, the `Newcomb-Nordlie fraternity' study \cite{nordlie1958,newcomb1961} recruited two waves of 17 male students who did not know each other, gave them free `fraternity-style' housing, and studied how their personal characteristics, political positions, and interests affected their eventual friendship formation. These two studies established that proximity plays an important role in friendship, but also that proximity is not sufficient, and that other characteristics matter. 

Later work \cite{latane1995} further quantified the relationship between distance and friendship, finding ``the inverse square of the distance separating two persons'' to be a good fit to measures of social impact. A retreat for all incoming sociology majors at the University of Groningen provided another opportunity for studying the emergence of friendship, with van Duijn et al. \cite{vanduijn2003} finding that friendships developed due to one of four main effects: physical proximity, visible similarity, invisible similarity, and network opportunity. 

There are also a number of studies using data from online social networks or other online platforms to study the connection between friendship and geography, although many of these are at the global scale \cite{liben2005,quercia2012,leskovec2014,cho2011,backstrom2010} and do not have resolution at the scales at which interactions can occur. Furthermore, users of location-based social networks, or those who share locations on general online social networks like Twitter and Facebook, are a fraction of total users and form an unrepresentative sample \cite{malik2015} for a general smartphone-using population. Other work has taken call logs and used calls as the ties to study with respect to geography \cite{onnela2011,wang2011}, although call logs and SMS have a surprisingly poor relationship with friendship as measured by self-report \cite{wiese2014,wiese2015}. 

A landmark series of papers by Bernard, Killworth, and Sailer \cite{killworth1976,killworth1979,bernard1977,bernard1979,bernard1982} showed that people are generally bad at recalling objective patterns of interaction. However, subsequent work \cite{freeman1987,krackhardt1987} argued that \emph{psychological perceptions of the network}, rather than objectively measurable ties of interaction, were causal for individuals' behavior: subjective data may in some cases be more valuable for predicting and explaining, and thus the psychological perceptions captured by survey data may be more valuable for certain tasks than objective measurements. 

This is related to adams' [sic] \cite{adams2010} critique of the model in Eagle et al. \cite{eagle2009} predicting friendship from co-location; adams notes that there are `close strangers and distant friends', both of which a method based on co-location would misclassify. In response, Eagle et al. \cite{eagle2010} argue that some causal network processes might happen unconsciously, and sensor measurements might be able to detect these. 
The correlation of friendship and proximity can make it difficult to sort out which may be causal for a given process \citep{cohen2008}. But in the future, if we can decorrelate friendships and proximity by controlling for one or the other, it could help us understand whether social influence or environmental factors are more likely to be the causal factor. Furthermore, for many mobile use cases, even imperfect friendship detection will be relevant, making the task of friendship detection from co-location a worthwhile pursuit. 

\subsection{Sensors for social networks}
Since the first work with the `sociometer' (later, `sociometric badge') in 2002 \cite{choudhury2002}, research groups have been using sensors to collect data about social interactions. Studies often used sensor `nodes' or other custom devices \cite{paradiso2010,angelopoulos2011,forster2012,friggeri2011,hsieh2010,laibowitz2006} and, with the exception of RFID tags \cite{barrat2013}, studies have largely used mobile devices \cite{stop2014,stop2015,kiukkonen2010,rachuri2011b,li2012,do2011,eagle2006,raento2009,kjaergaard2012,lawrence2006,sekara2014,sekara2014} because of their wide range of existing sensors, and because their use results in lower participant burden than when participants are required to wear or carry an additional device. The data used is either co-location data (from either GPS, self-reported check-ins, or mutual detection of fixed sensors or WiFi hotspots) or proximity (through Bluetooth or RFID). There are also studies that use video to detect interactions \cite{chen2007,ke2013,kong2016,vangemeren2016}, and sociometric badges \cite{choudhury2002} or headsets \cite{madan2006} that gather audio data that allows for detecting conversations between specific individuals, but, in our work, we focus on the co-location and proximity sensing made possible with mobile phones. 

These existing studies largely fall into three categories. Most common are studies that describe infrastructure, technical details, and study design, followed sometimes by descriptive modeling of network characteristics and some bivariate relationships \cite{stop2014,stop2015,sekara2014,sekara2014,kiukkonen2010,li2012,lepri2012,choudhury2003a,choudhury2003b,chronis2009,angelopoulos2011,aharony2011,barrat2008,cattuto2010,barrat2012,barrat2013,forster2012,friggeri2011,isella2011,miklas2007,rachuri2011b}. This has been important work that has contributed to present-day mobile sensing tools -- tools that are capable of going beyond exploration and overall descriptives into measuring specific processes, and applications designed on top of those processes like network-based health interventions \cite{madan2010b}. 

The second category is of those that try to build systems other than the specific sensing platform. They may build or lay the groundwork for recommender systems \cite{chin2012,chin2013,ganti2008,lawrence2006}, or present models or algorithms for mining information about interactions from sensor data \cite{do2011,do2013,eagle2006,phung2008}. 

The third category is of those that employ statistical models or techniques to make conclusions or predictions using sensor data. Stehl\'{e} et al. \cite{stehle2013} look at the connection between `spatial behavior', measured by RFID badges, and gender similarity. The `Friends and Family' or `SocialfMRI' study and dataset \cite{aharony2011} has been used to look at the connection between interaction and financial status \cite{pan2011} and interaction and sleep and mood \cite{moturu2011}. Madan et al. \cite{madan2011} used sensor measurements to look at the relationship between social interactions and changes in political opinion. The `Reality Mining' dataset \cite{eagle2006} has similarly been used to look at obesity and exercise in the presence of contact between people \cite{madan2010}. Another approach is that of Eagle \& Pentland \cite{eagle2009b}, which presents a spectral clustering system for extracting daily patterns from time series. Staiano et al. \cite{staiano2012} use ego networks (induced subgraphs of single nodes and all their respective neighbors) in call logs, Bluetooth-based proximity networks, and surveys to predict Big-5 personality traits. 

In this third category, and most similar to our work, are two papers based on the Reality Mining dataset \cite{dong2011,eagle2009}. Eagle and Pentland \cite{eagle2009} were the first to use mobile phone data proximity to infer a network of self-reported friendship ties. They first calculated a `probability of proximity' score over the range of a week as an average frequency of proximity over nine months of data, which they showed was systematically different for each of reciprocated self-reported ties $(A_{ij}, A_{ji}) = (1,1)$, non-reciprocated ties $(A_{ij}, A_{ji}) \in \{(0,1), (1,0)\}$, and no ties $(A_{ij}, A_{ji}) = (0,0)$. But their model did not use cross-validation, and their findings were based on aggregating over nine months of data to model a friendship self-report from the first month, which does not match a mobile use case which would involve detecting friendships only from recently gathered batches of location data. 

\begin{figure}
\includegraphics[scale=.5]{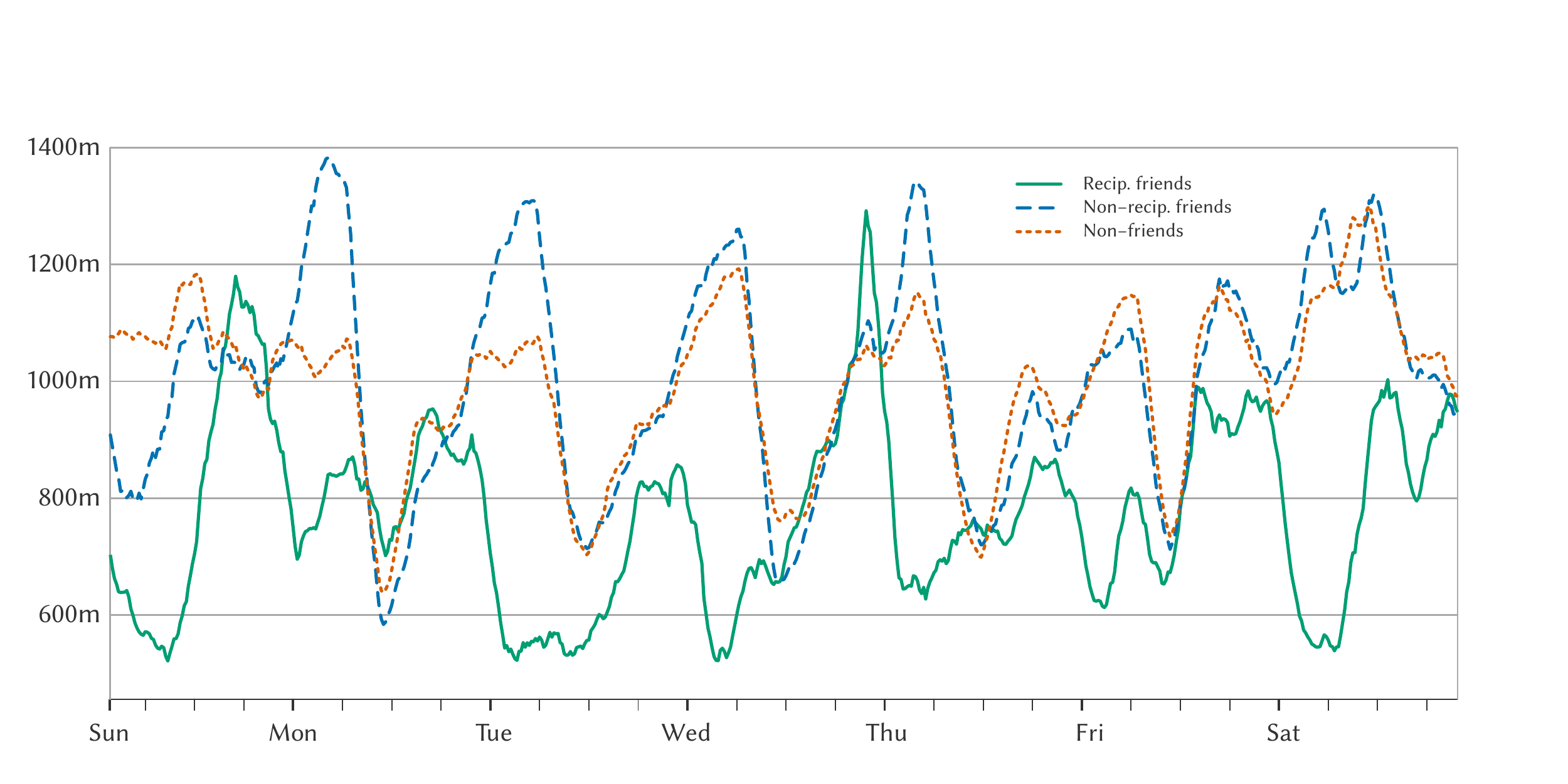}
\caption{The median weekly pairwise distances between reciprocated (mutual) friendships, $A_{ij} = A_{ji} = 1$, non-reciprocated friendships, $A_{ij}  = 1 \neq A_{ji} = 0$ or $A_{ij}  = 0 \neq A_{ji} = 1$, and non-friendships, $A_{ij} = A_{ji} = 0$, for times when pairs are within the area of the university, and aggregated over the entire period of data (i.e., no training/test split). This is analogous to the approach of \cite{eagle2009}, and this figure reproduces their figure 2 (except with median distance, rather than mean frequency of proximity). While it appears there is a strong pattern, it is a result of an aggregation that obscures the variance between weeks and in data splitting, such that this seeming pattern proved ineffective as a basis of classification in testing.}\label{fig:rm}
\end{figure}

We explored replicating their approach; while we also found that, when aggregated over the entire time period of data collection, there was a major difference between mutual friendship ties and both non-ties and non-reciprocated ties (fig. \ref{fig:rm}), this pattern proved ineffective for building a classifier because of how aggregation like this, over time, dyads, and splits of training and test sets, obscures the variance that poses challenges to good test performance. 

Dong et al. \cite{dong2011} also modeled the co-evolution of behavior and social relationships from mobile phone sensor data. They outlined a model that predicted self-reported friendships from sensor data (and other survey data), but also did not use cross-validation, and only reported one performance metric: that the binomial model explained 22\% of overall variance, of which 6\%  was due to sensor data. This presumably from a pseudo R-squared metric, but as the specific metric is not given and there was no cross-validated performance reported, it is difficult to compare results.

We now describe the questions we seek to answer and the study and analysis we conducted to answer them.

\section{Study design and procedure}
The goal of our study is to understand the feasibility of inferring social relationships (friendship in particular) from (only) passive smartphone data. We are especially interested in the following questions:

\begin{enumerate}
\item How well can we detect friendships from co-location features? In other words, if all we know about two people in a social system is their location patterns, how accurately can we say if they are friends?
\item If we know that friendships exist, how well can we detect if these  friendships are {\it close} friendships?
\item How accurately can we detect whether a friendship is likely to change? Will co-location patterns provide information about the creation or dissolution of friendships?
\end{enumerate}

To answer these questions, we carried out a 3-month study among members of a fraternity to use smartphone data to try and capture interactions and relationships as they were formed and evolved during that period. The following section describes the study setup and data collection process.

\subsection{Participants and recruitment}
We recruited members of an undergraduate fraternity in a research university in the northeastern United States. The fraternity had 60 members at the start of the study, with an additional 21 prospective members going through the `pledging' process during the study duration, of which 19 completed the process. Of this cohort of 79 men, we recruited 66 participants, of which 53 ultimately participated in sensor data collection, and of which 48 responded to at least one survey wave. Having this sort of well-defined {\it boundary specification} \cite{laumann1983} let us ask each study participant about their friendships with each member of the fraternity, giving negative examples that are explicit, unlike open-ended solicitation for friendships (such as from `name generator' instruments) in which individuals are only implicitly not friends by not being mentioned. 

The fraternity was relatively loose-knit; about 20 fraternity members live in a fraternity house, with the rest living elsewhere and required to be in the fraternity only one day a week (for a fraternity chapter-wide meeting). 
%
%
Participants were compensated \$20 a week for having the passive and automated sensor data collection software, AWARE \cite{ferreira2015}, installed on their smartphones, with additional \$5 incentives for each survey wave they completed. 

\subsection{Data collection}
Our task was to use mobile phone sensor data relating to location and proximity in a model that could recover self-reported friendship ties, and changes in such ties. Consequently, we collected survey data about friendships in three waves, and used AWARE to collect Wifi, Bluetooth, and location data from mobile phones.

\subsubsection{Survey data}
During the study, participants were asked to fill out a survey asking about their social connections, based off of existing instruments \cite{knecht2010,vandebunt1999}. There was a public listing of fraternity members, and consequently we were able to ask about respondents' ties to all fraternity members (i.e., ask about ties to everybody in the specified boundary), not just those participating in the study; while we were not able to relate friendships with non-participant fraternity members to sensor data, since non-participation meant we do not have sensor data, it does give a sense of the importance of non-participants in the social system.\footnote{For example, if non-participants were all seldom nominated by respondents (corresponding to low indegree in the collected networks), it would mean that study non-participation is related to being unimportant in the social system, which would be encouraging, although this did not turn out to be the case.} 
	
The surveys were collected three times over 9 weeks: shortly after the beginning of the study, then four weeks after, and lastly at the end of the study five weeks later (we made the second period longer, as one of these five weeks was spring break, when many study participants were away from campus). Participants were asked about five different quantities: their recollections about who they interacted with frequently; who they considered to be a friend; who they considered to be a {\it close} friend; who they went to for advice on personal matters; and who they went to for advice on professional/academic matters. The correlation between these collected networks, between each other and over time, is given below in figure (\ref{fig:jaccard}). Friendship can change at shorter intervals than six weeks; but since friendship is an internal and subjective psychological construct, currently the only way of getting data on friendship is surveys with high respondent burden that makes it infeasible to collect at more frequent intervals.
%
%
%

\subsubsection{Passive smartphone data}
We equipped each participant with the AWARE mobile phone framework\footnote{\url{http://www.awareframework.com}} \cite{ferreira2015} on their iOS devices ($\approx$90\% of participants) or Android devices (the remaining $\approx$10\%). There were no users of Windows or other mobile operating systems. We used AWARE to record Bluetooth and WiFi detections, each at 10 minute sampling intervals. We also had continuous monitoring of battery and screen status (on/off), and complete records of call and message metadata (with hashed values for phone numbers). 
For location, the Android AWARE client uses the Google fused location plugin, which has several options for trading off accuracy and battery usage, and for which we selected the low power option. The iOS AWARE client uses the iOS location services, in which we similarly selected an option with low battery usage. 

We also performed WiFi fingerprinting in the fraternity house to help us determine when participants were co-located in rooms in the house. 

\section{Data processing}
\subsection{Data Handling}
\subsubsection{Survey data}
\begin{figure}
\includegraphics[scale=.5]{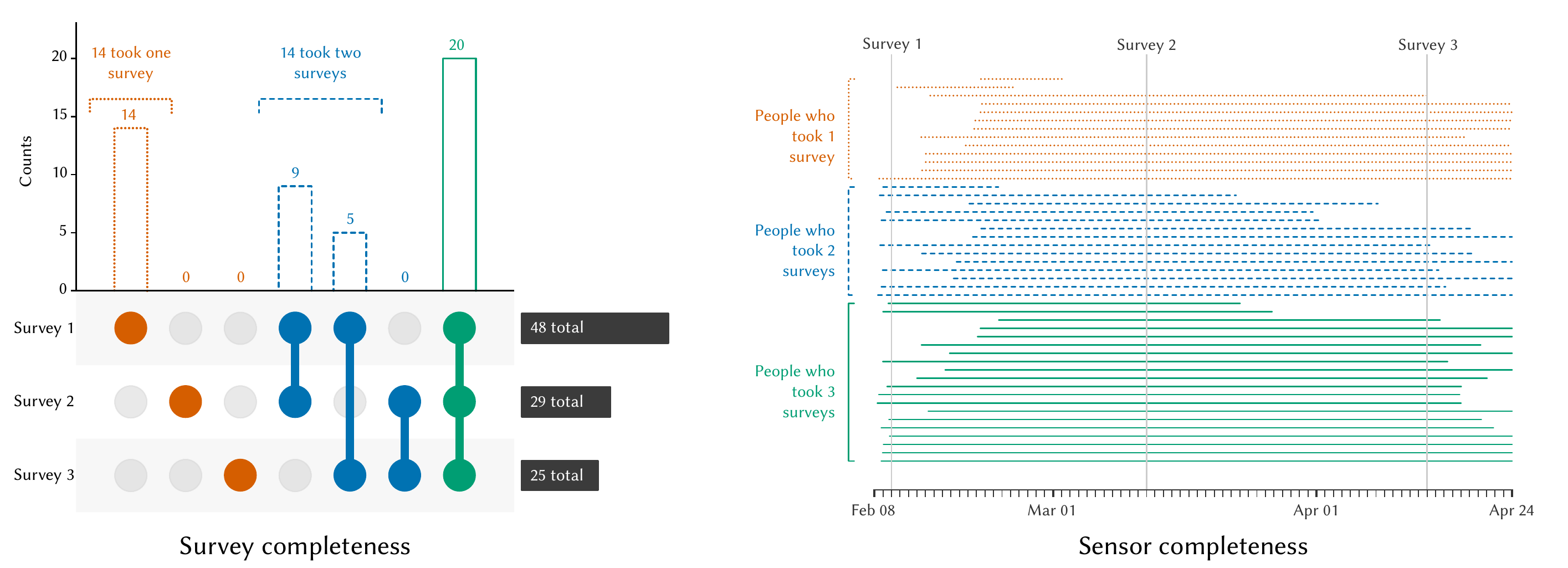}
\caption{(a. Left) Looking at longitudinal completeness, 14 people completed survey 1 only, and none completed surveys 2 or 3 only. 9 people completed surveys 1 and 2, 5 people completed surveys 1 and 3, none completed 2 and 3 only, and 20 people completed all three waves. This is shown in the vertical bars at the top. This comes out to 48 respondents for survey 1, 29 respondents for survey 2, and 25 respondents for survey 3, shown in the solid horizontal bars on the right side. (b. Right) We show the time periods in which sensor data was collected for people  who answered one survey (dotted lines), two surveys (dashed lines), or all three surveys (solid lines). The times of the three surveys are marked with vertical lines.}
\label{fig:completeness}
\end{figure}

The completeness of the survey data is shown in figure (\ref{fig:completeness}a). The response rate dropped in each survey round; compared to survey 1, survey 2 had a response rate of 59\%, and survey 3 had a response rate of 51\%. In total, there were 48 participants providing network data, 34 of which responded to 2 surveys giving us longitudinal network data (the minimum requirement for detecting changes in friendship), including 20 participants that responded to all 3 surveys. In total, out of $\binom{48}{2} = 1128$ potential pairs, we were able to train and/or test on $830$ pairs.

\subsubsection{Sensor data}
The completeness of the sensor data is shown in figure (\ref{fig:completeness}b). Some logistical problems prevented all participants from starting smartphone data collection on the first day, and some participants discontinued the use of the app because of technical issues (battery life, sporadic interference with certain external Bluetooth devices, etc.). 

There were two sources of missing values in the calculated features: either artifacts relating to no observations fulfilling a certain criteria (e.g., no co-locations within 50m on mornings), or else actual missing data (one or both mobile devices were not providing a certain sensor's data during a given period, e.g., mornings of a given week). For the former (artifacts), we replaced missing values with appropriate substitutes, such as 0s or the maximum possible value. For logarithmic features, some of which could be less than 1, we replaced $-\infty$ with zeros. For inverse-squared features, we replaced $\infty$ with a value, 200, slightly larger than the largest observed inverse-squared value. 

\begin{figure}
\includegraphics[scale=.5]{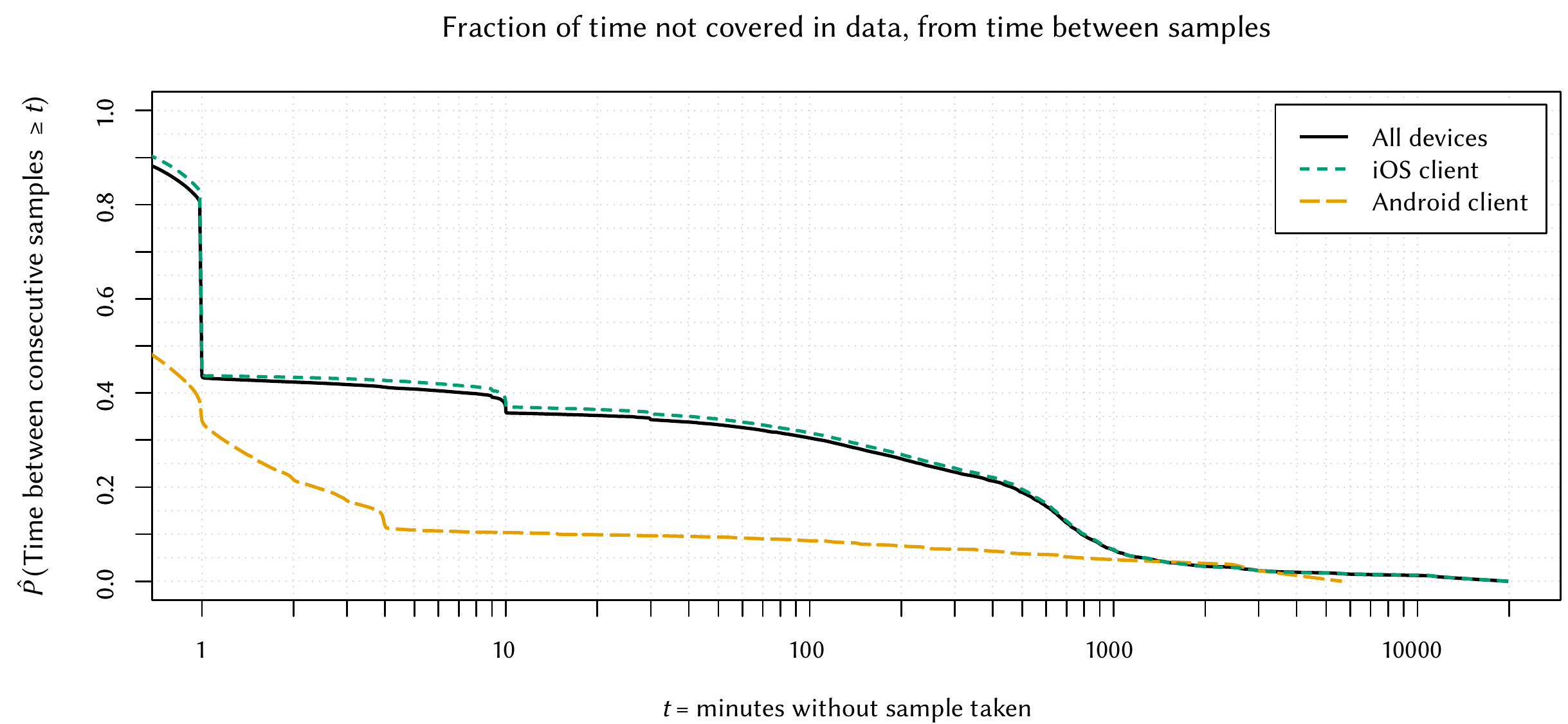}
\caption{A plot of the survival function (the empirical complementary cumulative distribution) for time not covered by sensor data. 40\% of the study time across all participants is covered by sensor readings less frequent than 10 minutes, meaning that 60\% of the data has no gaps in coverage.}\label{fig:missing}
%
%
\end{figure}

Based on the distribution of lengths of time where data was missing (fig. \ref{fig:missing}), we would have needed to interpolate up to eight hour intervals to have any real impact on the proportion of time that has missing data. Consequently, we chose to not use partial interpolation, and kept cells of missing values in the feature matrix. This necessitated using classifiers that can handle missing values among the features, like the \textsf{R} random forest implementation \texttt{rpart} \cite{therneau2018} which has procedures for handling missing values when constructing decision trees, and other packages built on top of \texttt{rpart}. 

We did test our assumptions about the importance of maintaining missing values by trying different variations. We did try out last value carried forward interpolation on the time series prior to feature extraction, as well as mean, median, and mode interpolation on the matrix of extracted features, but neither improved results. 

\subsubsection{Spring break}
Spring break may be extremely informative, for example if two people are proximate to each other but far from everybody else it may be that they are more likely to be friends. However, spring break is systematically different from every other week, such that if we train on spring break, we have no meaningful test set. Thus, we removed spring break from the data set. This is also why the two periods have  an unequal number of weeks, with 4 weeks  between survey waves 1 and 2, and 5 weeks between survey waves 2 and 3; spring break fell between survey waves 2 and 3, such that removing it leaves 4  weeks in each period. 


\subsection{Collected surveys and sensors}
\subsubsection{Network survey instrument}
We asked participants about five different types of ties in each of the three surveys: following previous social science literature, we asked about advice-seeking relationships (both personal advice seeking, and academic/professional advice-seeking), in addition to asking about friendships and, for each reported friendship, asking if it was also a close friendship. For comparison with work on recall \cite{bernard1977,bernard1979,bernard1982,killworth1976,killworth1979} and memorability of social interactions \cite{latane1995}, we also asked about frequency of interaction. 

\begin{figure}[ht]
\includegraphics[scale=.75]{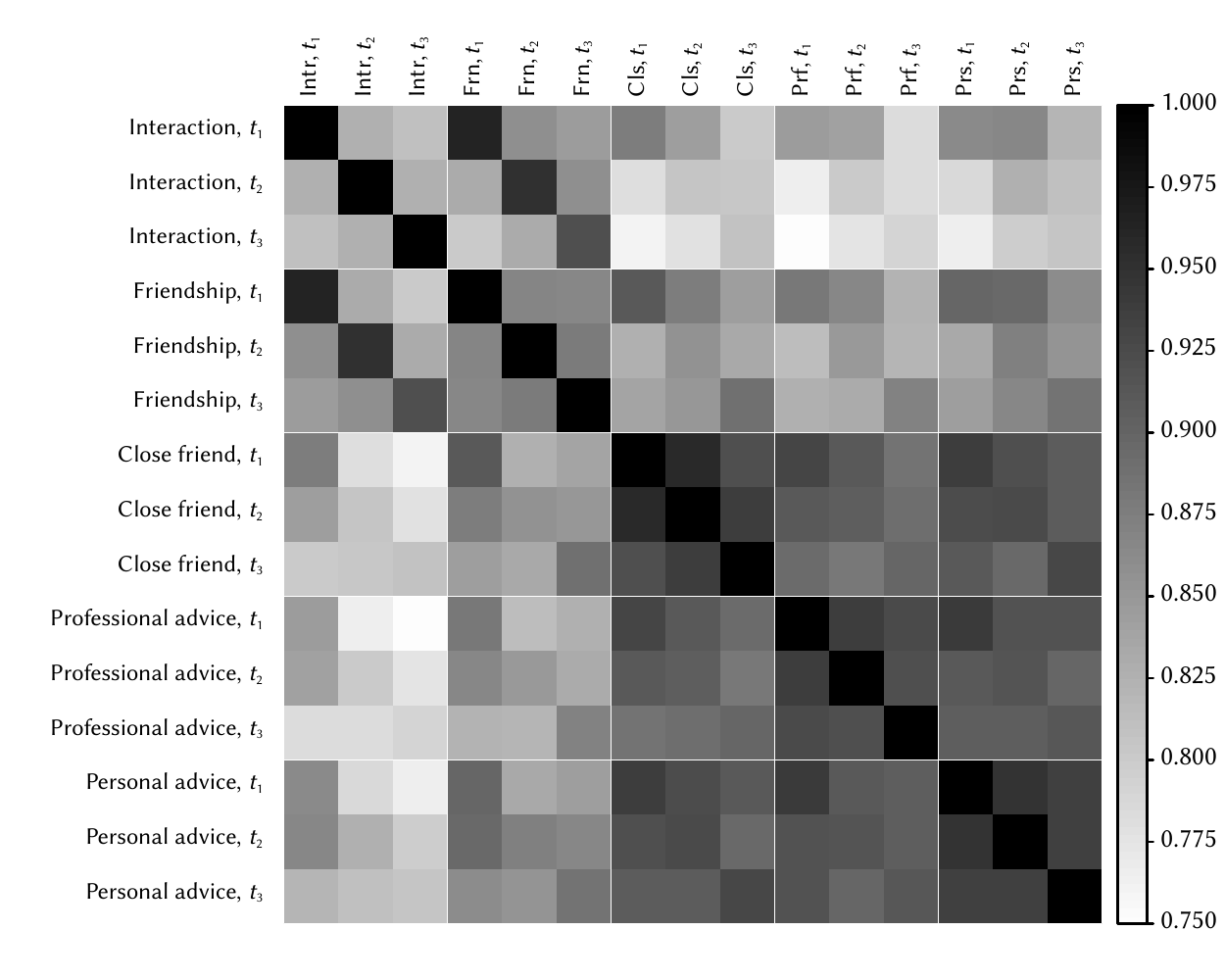}
\caption{The similarity between networks (5 types of ties, each collected 3 times), measured via the Jaccard index. The self-reported frequent interaction and friendship networks are more similar than the other networks, and both also exhibit more variation across the three waves.}\label{fig:jaccard}
\end{figure}

The similarities between these collected networks, both for the same network across the three waves and between the different networks, is given in figure (\ref{fig:jaccard}). The similarity metric used is the Jaccard index, a common method for comparing networks (as it looks only at ties shared across the two networks, not shared non-ties), potentially of overlapping but unequal sets of nodes, which in our case happens because of non-response. For two networks $N_A$ and $N_B$, with $n_{A \cap B}$ overlapping nodes and adjacency matrices $A$ and $B$ restricted to these nodes, the Jaccard index is 
\[
J(A, B) = \frac{| \{A_{ij} = B_{ij} = 1\} |}{2 \times \binom{n_{A \cap B}}{2}}
\]

As we can see, there is a much higher correlation between self-reported frequent interaction and friendship than there is between friendship and close friendship. There is also a high correlation between close friendship and the two types of advice ties; while we did not use advice ties in the current analysis, this similarity gives insight into what types of relationships the prompt about `close friendships' elicit (like the prompt about friendship, we explicitly do not define what we mean by `close', letting participants interpret the term). 

Looking at the changes in the networks from survey to survey, we see that close friendships and both types of advice-seeking relationships are much less variable over time than are friendships or self-reports of frequent interaction. 

\subsubsection{Bluetooth}
Collected Bluetooth data turned out to be unusable. Both Android and iOS no longer make the 16 hex digit Bluetooth MAC addresses of detected devices available to app developers. Instead, detected devices are recorded in terms of a 32 hex digit universally unique identifier (UUID), which are assigned by the detecting device uniquely to each detected device and used to recognize those detected devices in the future. 
%

\subsubsection{Call logs and SMS}
Following the findings of Wiese et al. \cite{wiese2014,wiese2015} that call logs and SMS do not necessarily help detect degree of friendships, we elected to restrict our attention to using co-location only. Another reason to  avoid call logs and SMS are that such communications metadata are already seen as sensitive and intrusive, even if they were to turn out to not help us predict our target of interest. Lastly, there was a substantive reason to not use communications data: we were informed that the fraternity largely used a group chat application for communications with one another, such that we expected call logs and SMS to not capture any informative aspects of communications. 

\subsubsection{Wifi}
One candidate for characterizing proximity is when two devices detect or connect to the same Wifi device. Here, Wifi hotspot MAC addresses are unique (unlike the hotspot name/label, which for example with `eduroam' is shared not only across multiple hotspots in the same university, but across multiple cities across the world!), and mutual detection of this picks up when two devices are proximate. 
Out of 830 potential pairs, 406 pairs of mobile devices detected at least one Wifi hotspot in common (although not necessarily at the same time). 

As mentioned above, we also conduced Wifi fingerprinting in the fraternity house, including collecting all Wifi devices detected in each room along with the received signal strength indication (RSSI) of the respective signals. In order to perform Wifi fingerprinting (match a set of hotspots that are detected by a mobile phone in a given scan and with respective RSSIs to previously collected profiles from specific rooms), we needed multiple detected hotspots per scan (every 10 minutes). However, we found that only about 6.7\% of scans for Wifi hotspots recorded more than one detected hotspot; in the frat house as well, we could tell when a device was connected to one of the frat house's Wifi hotspots, but not which other hotspots were detected in order to determine a specific room. Thus, we only use as the basis for features whether at least one Wifi hotspot was detected in common at the scan of a specific 10 minute interval from two devices, ignoring the tiny fraction of detections that include multiple devices, and also ignoring RSSI. The frat house has 5 main Wifi devices for about 30 rooms over 3 floors. Based on the size of rooms in the fraternity house and the relative coverage of its Wifi devices, we estimate that at least within the fraternity house, our Wifi localization approach is accurate to within a bit of a smaller radius than its general 32m accuracy, perhaps 20m or so; however, we do not have similar measurements for the rest of campus. 



\subsubsection{Location}
Since we have, from previous theory, that co-location is causally related to friendship through interaction, we would ideally want to extract features from pairwise distance measurements that will be effective as a proxy for interaction. However, in the Google Fused Location plugin that AWARE uses to collect location data, we used the PRIORITY\_LOW\_POWER option which prioritizes low power usage, as previous testing with AWARE had showed battery drain was a major cause of participant dropout. This low power option does not actively use GPS, instead using a combination of cell phone towers and detected Wifi hotspot with known geolocations, and is advertised as being accurate to within about 10km.\footnote{\url{https://developers.google.com/android/reference/com/google/android/gms/location/LocationRequest}, and \url{http://www.awareframework.com/plugin/?package=com.aware.plugin.google.fused_location}} 
In the iOS client, the accuracy setting corresponding to low power use was to set \texttt{desiredAccuracy} option to 1km, with a threshhold for recording new movements of 1000m.\footnote{\url{https://developer.apple.com/library/content/documentation/UserExperience/Conceptual/LocationAwarenessPG/CoreLocation/CoreLocation.html}} In practice, the reported accuracy was usually much better, with a significant portion of readings reporting an accuracy of within 10m. 

Additionally, when calculating the continuous-valued time series of pairwise distances, we also generated binary time series for if the locations of both members of the pair fell within a geobox around the university's campus, and a geobox around the fraternity house.

\begin{figure}
\includegraphics[scale=.25]{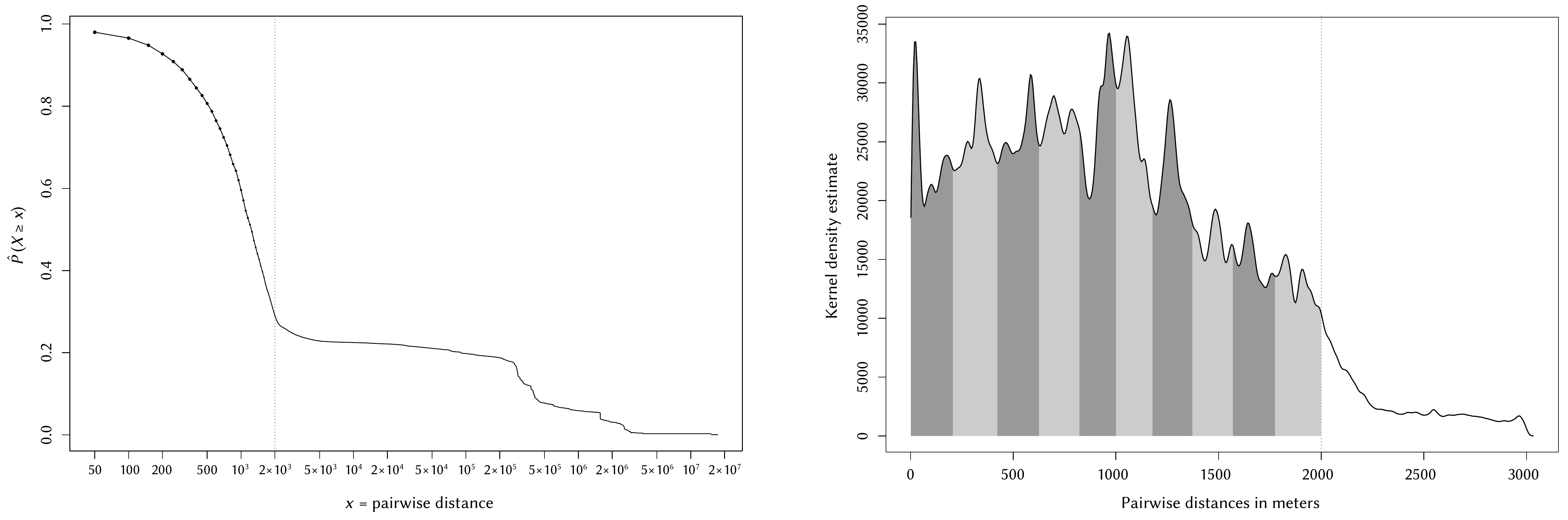}
\caption{A survival function (a. left), plotted in log-$x$ scale, shows pairwise distances over time. Based on the `elbow' around 2000m (approximately the size of the university and surrounding area), marked with a vertical dotted line, we only found clusters for pairwise distances below 2000m. Below 2000m, we clustered distances (again weighted by the time spent at that distance). The fitted clusters are shown on top of a kernel density estimate (b. right) that gives a detail of the head of the distribution. The cluster breaks are at 207m, 422m, 626m, 822m, 1001m, 1178m, 1373m, 1570m, 1776m, and then our cutoff of 2000m. These are also listed in table (\ref{tab:features}).}\label{fig:pairwise}
\end{figure}

As a way of reducing the continuous-valued time series, we sought to pick several choice thresholds that might characterize geographic similarity in simple way. First, we plot an empirical complementary cumulative distribution function (i.e., a survival function) in log-$x$ scale (fig. \ref{fig:pairwise}a) to see the overall distribution. There is an `elbow' around 2000m, which is about the size of the university and surrounding area. Then, within 2000m, we use 1-dimensional clustering \cite{wang2011}, weighted by time and using 10 clusters, and used the boundaries of the fitted clusters as thresholds. These thresholds are shown as the boundaries regions of gray over a kernel density estimate of the distribution over the first 2000m (fig. \ref{fig:pairwise}b). 


After data processing, we have the following:
\begin{itemize}
\item 1 continuous-valued time series of {\it pairwise distances}
\item 10 binary time series of whether both members of a given pair were within a given threshold of each other
\item 1 binary time series of whether both members of a given pair were within a geobox around the university campus
\item 1 binary time series of whether both members of a given pair were within a geobox around the fraternity house
\item 1 binary time series of whether both members of a given pair detected at least one Wifi hotspot in common
\item 1 binary time series of whether both members of a given pair detected a Wifi hotspot visible from the fraternity house in common
\end{itemize}
Next, to compare self-reported ties and co-location, it is necessary to summarize these time series into a set of features. These are summarized in table (\ref{tab:features}).

\begin{table}
\includegraphics[scale=.825]{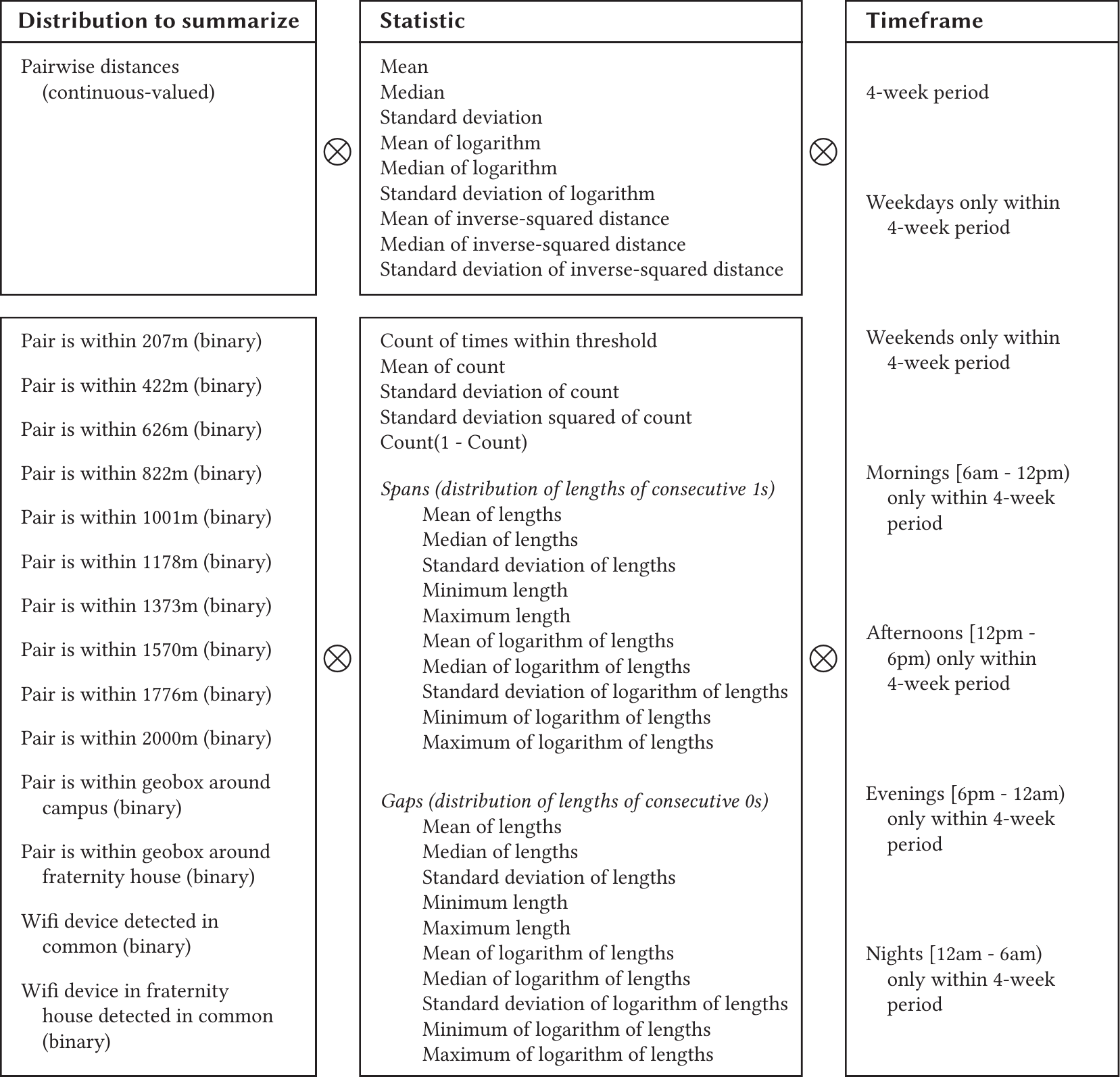}
\caption{Extracted features. ``$\otimes$'' indicates taking all pairwise combinations. The thresholds are irregularly spaced because they are empirically derived from 1-dimensional clustering; see figure (\ref{fig:pairwise}b) for these clusters.}\label{tab:features}
\end{table}

For the continuous time series and each of the binary time series, we extract relevant summary  statistics relating to central tendency, variance, and (if applicable) range. We employed summaries of distributions after logarithmic transformation after observing that the original distributions were often heavily right-skewed. Additionally, for the binary time series, we can consider the length of sequences of consecutive 1s ({\it spans} of co-location at the given threshold) and of consecutive 0s ({\it gaps} between co-location at the given threshold). These are integer-valued but we treat them as continuous, and calculate an additional set of summary statistics accordingly. All of these summary statistics are given in the central column of table (\ref{tab:features}). 

Lastly, each of these feature types are crossed with time periods: weekdays only and weekends only, nights only (12am - 6am), mornings only (6am - 12pm), afternoons only (12pm - 6pm), and evenings only (6pm - 12am). These are shown in the right column of table (\ref{tab:features}). 

In total, there are $9$ features for the continuous-valued time series of pairwise distances, $12 \times (5+20) = 300$ features for the binary time series, and each of these 309 features are calculated over seven settings, for $309 \times 7 = 2163$ candidate location features. 
Wifi features were $2 \times (5+20) = 50$, and $50 \times 7 = 350$ for an additional 350 features, for a total of 2,513 features. We extracted these over two 4-week periods, corresponding to the 4 weeks between surveys 1 and 2, and the 5 weeks between surveys 2 and 3 with the week of spring break subtracted out.

\section{Modeling targets and evaluation methods}

\subsection{Modeling targets}
We take on three targets for modeling.
\begin{enumerate}
\item Detecting friendship. This is a standard binary classification task. In this task, we do not make use of survey wave 1.
\item Detecting friendship strength. Given that two people are friends, can we detect whether or not they have reported that the friendship is a close one? For this, we restrict the data set to instances of friendship ties only, and do binary classification of close friendships. Again, we do not make use of  survey wave 1.
\item Detecting change in friendship. Here, our targets are \begin{itemize}
	\item $P(A_{ij}^{(t)} = A_{ij}^{(t+1)} \mid X^{[t, t+1)})$: No change in friendship (either no friendship, or maintained friendship)
	\item $P(A_{ij}^{(t)} \neq A_{ij}^{(t+1)} \mid X^{[t, t+1)})$: Change in friendship (either tie creation or tie dissolution)
\end{itemize}
While we ideally would be able to separately model tie creation and dissolution, as they are distinct processes \cite{snijders2010}, in our data only a small proportion of ties changed in either direction such that modeling became difficult. We will see below that this modeling target was the most challenging of all, although treating it as a multiclass problem over the direction of change only led to worse performance.
\end{enumerate}

\subsection{Cross-validation schema}
In order to comprehensively evaluate our classifier's performance at detecting friendship, we use three cross-validation schema. Each corresponds to a different use case, and tests the generalizability of our method to that use case. In each case, dependencies (redundancies in data, latent or unmodeled similarities) between training and test sets can share information across a split in data,  dependencies that would not be present in application settings, therefore inflating test performance compared to real-world performance. 

Each schema uses a different rule and use case to  assign observations to training and test folds. The rules and use cases of these schema are detailed below.  

\subsubsection{Cross validation with unrestricted assignment}
This is independently assigning each observed $A_{ij}$ to a fold. It corresponds to a use case where a model is trained on a population ($n-k$ pairs) and then applied back to $k$ pairs from same population (potentially seeing the same people multiple times, or the same dyad in multiple directions).

\subsubsection{Cross validation with dyadic assignment} 
This groups all values associated with a pair of individuals (a dyad), that is, $(A_{ij}^{(1)}, A_{ji}^{(1)}, A_{ij}^{(2)}, A_{ji}^{(2)}, A_{ij}^{(3)}, A_{ji}^{(3)})$, and assign the entire 6-tuple to a single fold. Some values in the tuple will be missing, causing folds to be of different sizes; But since assignment to fold is not dependent of the number of missing values, sizes will be the same in expectation. 

Such assignment controls for reciprocity and temporal autocorrelation. For reciprocity, if $A_{ij} = A_{ji}$, then the label-feature pair $(A_{ij}, X_{ij})$ and $(A_{ji}, X_{ji})$ are identical and should not be split between training and test. Similarly for temporal autocorrelation, if two people's friendship and co-location patterns do not change over time, then $(A_{ij}^{(t)}, X_{ij}^{[t-1, t)})$ and $(A_{ij}^{(t+1)}, X_{ij}^{[t, t+1)})$ would also be very similar and should not be  split between training and test. 

Cross validation with dyadic assignment corresponds to a use case where we have not previously seen the labeled co-location patterns of a given dyad, whether previously in time or in one direction, to have included it as a training instance. 

\subsubsection{Cross validation with temporal block assignment}
This splits data by whether a class label is from survey 2 or survey 3 (for detecting friendship and strength of friendship) or is the change from survey 1 to 2 or the change from survey 2 to 3 (for detecting change in friendship). In other words, for detecting friendship and strength, we  train on $(A^{(2)}, X^{[1, 2)})$ and test on $(A^{(3)}, X^{[2, 3)})$, and for detecting change, we train on $(A^{(1)}, A^{(2)}, X^{[1, 2)})$ and test on $(A^{(2)}, A^{(3)}, X^{[2, 3)})$. 

As a note, here we can only split into 2 folds as we only have two observation spans between different surveys. 
Cross validation with temporal block \cite{bergmeir2012,racine2000} assignment accounts for temporal variation in co-location. If there is a great deal of variability in co-location patterns, then our classifier would have little generalizability over time. In this case, if we train with instances with features from both $X^{[t-1, t)}$ and $X^{[t, t+1)}$, it would even out the temporal variation and obscure the lack of generalizability. But if we train only on instances associated with features $X^{[t-1, t)}$ and then test only on instances associated  with features $X^{[t, t+1)}$, it simulates how well out classifier will do in predicting friendships from future patterns of co-location data. 

\subsection{Evaluation metric}
To summarize classifier performance, we rely on the Matthews correlation coefficient (MCC). This is the same as Pearson's $\phi$, or mean square contingency coefficient, an analog for a pair of binary variables of Pearson's product-moment correlation coefficient, but was rediscovered by Matthews \cite{matthews1975} for use as a classification metric. For the count of true positives (TP), true negatives (TN), false positives (FP) and false negatives (FN), the MCC is
\[
\text{MCC} = \frac{TP \times TN - FP \times FN}{\sqrt{(FP + TP) \times (TP + FN) \times (TN + FP) \times (TN + FN)}}.
\]
The MCC has several desirable properties. First, like the F1 score and area under the ROC curve (AUC), it summarizes the performance on both classes in a single number. Unlike AUC and F1, however, it has an interpretable range: 0 for random predictions, -1 for perfect misclassification, and 1 for perfect classification. Most helpfully, is a good summary of performance in cases of class imbalance \cite{boughorbel2017}, which have here (about a 25:75 split). We include other metrics, but rely on the MCC as the single-number summary of how far we are above a random baseline of MCC = 0. Note that, if we predict the majority class for all instances, the MCC is also zero. 



%
\subsection{Feature Selection}
Feature selection can often improve classifier performance, but it is also useful for diagnostic and exploratory analysis. In our case, we are interested in a reduced set of features that can provide similar or better classification results, and that may be less burdensome to extract for use in real-time mobile applications built on friendship detection. To produce a selected set of features, we use Correlation-based Feature Selection (CFS) \cite{hall1999}, which selects features that are both correlated with the class label, and uncorrelated with one another. 

To select the most stable set of features, we run the CFS method on the training set built with what turns out to be our most conservative cross validation scheme, temporal block assignment. We take the half of data with features extracted from the first four weeks and further divide it into 10 folds. We perform CFS of each fold, then look at the features that were selected in the maximum number of folds, an approach also applied  more formally elsewhere \cite{meinshausen2010}.

We choose those features that appeared in CFS runs on at least 9 of the 10 folds. These features are then entered in the classification process for friendship detection.

\section{Results}
\subsection{Friendship detection}
Results for the three cross-validation schemes are given in table (\ref{tab:perf1}). In each case, the no information rate corresponds to the proportion of the majority class, 0, and would be the accuracy we would get if we always predicted no tie. 

The unrestricted assignment gives better results than either of the other two CV schema, showing that labeling a previously unseen dyad is indeed a more specific and difficult task than what is evaluated by unrestricted assignment,  and that there is a significant amount of variation in co-location patterns over time---and that while our classifier performance does drop, it still generalizes across patterns in time. 

We use a one-sided binomial test of the accuracy against the No Information Rate (NIR), equal to the frequency of the majority class, and find that both unrestricted and dyadic CV are significant at the usual $p<0.05$ level. Under temporal block CV, the classifier is only significantly better than the NIR at the $p<0.1$ level.

In our classifications, the MCC ranges from .30 in CV with unrestricted assignment, to .26 in CV with dyadic assignment, and .21 in CV with temporal block assignment. This indicates that the classifier performance is between 30\% and 21\% better than baseline (for which MCC=$0$). 


%
%
%
%

\begin{table}[ht]
\centering
\resizebox{.6\textwidth}{!}{\begin{tabular}{|r|ccc|}
\hline
{\bf Cross validation} & {\bf Unrestricted} & {\bf Dyadic} & {\bf Temporal block} \\ \hline
Accuracy									&0.8006	&0.7920	&0.7913 \\
Accuracy, 95\% CI							&(0.7882, 0.8125)	&(0.7794, 0.8042)	&(0.7726, 0.8091) \\
{\it (No Information Rate / Majority class)}						&{\it (0.7740)}	&{\it (0.7740)}	&{\it (0.7785)} \\
Binomial test, Accuracy vs. NIR, $p$-value						&$p$=1.5e-05&$p$=0.0025	&$p$=0.0901	 \\ \hline
Precision (Positive predictive value)		&0.6918	&0.6508	&0.6812	 \\
Recall/Sensitivity (True positive rate)		&0.2122	&0.1723	&0.1088 \\
Specificity (True negative rate)			&0.9724	&0.9730	&0.9855 \\
F1 score									&0.3248	&0.2724	&0.2964 \\
AUC											&0.7148	&0.7039	&0.1876 \\
Matthews correlation coefficient			&0.3039	&0.2562	&0.2120 \\ \hline
\end{tabular}
}
\caption{Friendship detection, test performance across the three CV schema. The no information rate corresponds to a baseline accuracy given by predicting no ties; in the case of networks, this is 1 minus the density of the network.}\label{tab:perf1}
\end{table}

\subsection{Detecting close friendships}
We repeat the assessment of the above models, conditioning on the presence of a friendship, and making our detection target whether or not a friendship is reported to be {\it close}. In this case, the network of close friendships has a network density of .41, making the no information rate .59. 

\begin{table}[ht]
\centering
\resizebox{.6\textwidth}{!}{\begin{tabular}{|r|ccc|}
\hline
{\bf Cross validation} & {\bf Unrestricted} & {\bf Dyadic} & {\bf Temporal block} \\ \hline
Accuracy									&0.6817	&0.6670	&0.5741 \\
Accuracy, 95\% CI							&(0.6511, 0.7112)	&(0.6361, 0.6969)	&(0.5259, 0.6212) \\
{\it (No Information Rate / Majority class)}&{\it (0.5861)}	&{\it (0.5861)}	&{\it (0.5185)} \\
Binomial test, Accuracy vs. NIR, $p$-value	&$p$=7.6e-10&$p$=1.8e-07	&$p$=0.0117	 \\ \hline
Precision (Positive predictive value)		&0.6904	&0.6711	&0.7069	 \\
Recall/Sensitivity (True positive rate)		&0.4188	&0.3832	&0.1971 \\
Specificity (True negative rate)			&0.8674	&0.8674	&0.9241 \\
F1 score									&0.5213	&0.4879	&0.3083 \\
AUC											&0.6997	&0.6695	&0.5889 \\
Matthews correlation coefficient			&0.3250	&0.2906	&0.1777 \\ \hline
\end{tabular}
}
\caption{{\it Close} friendship detection, conditioned on the presence of a friendship, test performance across the three CV schema.}\label{tab:perf2}
\end{table}

We see a similar pattern of performance, with temporal block CV being the most conservative (18\% better than baseline), and unrestricted CV being more optimistic (32\% better than baseline). 

\subsection{Detecting changes in friendship}
Detecting loss in friendships could be particularly important for social interventions, such as preventing the onset of isolation. However, the rarity of changes in friendship (only 13\% of ties change, either being created or dissolving) complicates modeling. 

Our approach in meaningfully detect changes in friendship proved to be challenging. AdaBoost failed to predict any positive test cases for any CV schema; a random forest performed better with a Matthews correlation coefficient of .07 for the unrestricted CV and .03 for the dyadic-based CV (see table (\ref{tab:perf3}). The classifier output does not pass a statistical test for being significantly better than the No Information Rate. One of the reasons for the poor performance may be the type of features used in the classification. We used the same aggregated features used for friendship detection to detect change. However, change in friendship may be reflected in the feature values and thus a feature set that contains change values may better capture change in friendship.   

\begin{table}[ht]
\centering
\resizebox{.6\textwidth}{!}{\begin{tabular}{|r|cc|}
\hline
{\bf Cross validation} & {\bf Unrestricted} & {\bf Dyadic} \\ \hline
Accuracy									&0.6842	&0.8645	\\
Accuracy, 95\% CI							&(0.6692, 0.6989)	&(0.8532, 0.8752)\\
{\it (No Information Rate / Majority class)}&{\it (0.8710)}	&{\it (0.8710)}	 \\
Binomial test, Accuracy vs. NIR, $p$-value	&$p$=1&$p$=0.8902	 \\ \hline
Precision (Positive predictive value)		&0.1676	&0.2093	 \\
Recall/Sensitivity (True positive rate)		&0.3651	&0.0183	\\
Specificity (True negative rate)			&0.7315	&0.9898	\\
F1 score									&0.2297	&0.0336	\\
AUC											&0.5483	&0.5167	\\
Matthews correlation coefficient			&0.0720	&0.0256	\\ \hline
\end{tabular}
}
\caption{{\it Change} detection, random forest test performance. AdaBoost made only negative test classifications, but random forests (performance shown here) did make some positive classifications under unrestricted and dyad-based CV, although under temporal block CV again there were no positive classifications.}\label{tab:perf3}
\end{table}

\begin{figure}[h]
\includegraphics[scale=.3]{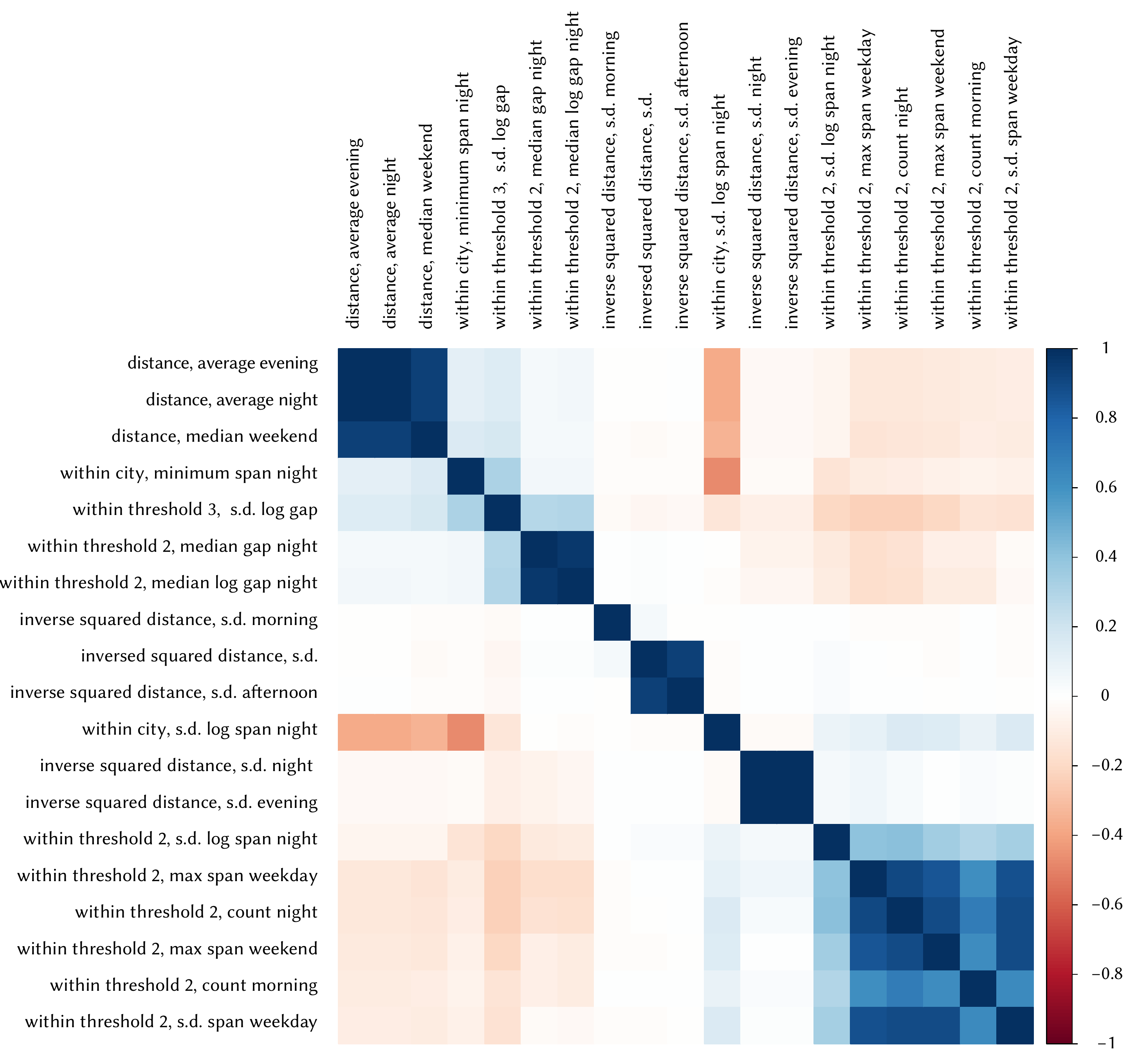}
\caption{Correlations between the features selected via CFS on the training set of a temporal block cross-validation scheme. The ordering is from the angular order of eigenvectors.}\label{fig:cormat}
\end{figure}

\begin{table}[h]
\centering
\resizebox{.6\textwidth}{!}{
\begin{tabular}{|r|lll|}
\hline
\textbf{Feature}                            & \textbf{Distribution}    & \textbf{Summary statistic} & \textbf{Timeframe} \\ \hline
1.  & Distance                 & Mean                       & Evening            \\
2.  & Distance                 & Mean                       & Night              \\
3.  & Distance                 & Median                     & Weekend            \\
4.  & Within city              & Minimum span               & Night              \\
5.  & Within threshold 3       & Log gap                    & All                \\
6.  & Within threshold 2       & Median gap                 & Night              \\
7.  & Within threshold 2       & Median log gap             & Night              \\
8.  & Inverse squared distance & S.D.                       & Morning            \\
9.  & Inverse squared distance & S.D.                       & All                \\
10. & Inverse squared distance & S.D.                       & Afternoon          \\
11. & Within city              & S.D. log span              & Night              \\
12. & Inverse squared distance & Standard deviation         & Night              \\
13. & Inverse squared distance & Standard deviation         & Evening            \\
14. & Within threshold 2       & S.D. log span                & Night              \\
15. & Within threshold 2       & Max span                   & Night              \\
16. & Within threshold 2       & Count                      & Night              \\
17. & Within threshold 2       & Max span                   & Weekend            \\
18. & Within threshold 2       & Count                      & Morning            \\
19. & Within threshold 2       & S.D. span                  & Weekday            \\ \hline
\end{tabular}
}
\caption{The 19 features selected via CFS on the training set from temporal block assignment: what they measure, how they summarize it, and the timeframe in which they summarize  it. Ordering is from angular order of eigenvectors on the correlation matrix (fig. \ref{fig:cormat}).}\label{tab:selected}
\end{table}

\subsection{Feature Selection}
While we applied CFS to select features from the  training set in all tasks, the features selected were not always consistent across folds, and across cross validation schema. So, we focus on the features selected in the case of the most conservative cross validation schema, and the  extent to which feature selection improved model performance here. 

Applying CFS to only the training data from temporal block assignment and splitting it into 10 folds, we find 19 features that are selected in 9 or 10 of the folds. Using only these features leads to improved test performance from temporal block assignment, shown in table (\ref{tab:perf4}), which also includes the test performance with this set of features under each cross validation scheme.

\begin{table}[ht]
\centering
\resizebox{.6\textwidth}{!}{\begin{tabular}{|r|ccc|}
\hline
{\bf CV assignment method} & {\bf Unrestricted} & {\bf Dyadic} & {\bf Temporal block} \\ \hline
Accuracy									&0.7975	&0.793	&0.7923 \\
Accuracy, 95\% CI							&(0.785, 0.8095)	&(0.7804, 0.8051)	&(0.7736, 0.8101) \\
{\it (No Information Rate / Majority class)}&{\it (0.774)}	&{\it (0.774)}	&{\it (0.7785)} \\
Binomial test, Accuracy vs. NIR, $p$-value	&$p$=0.0001&$p$=0.0016	&$p$=0.0734	 \\ \hline
Precision (Positive predictive value)		&0.6602	&0.6370	&0.5799 \\
Recall/Sensitivity (True positive rate)		&0.2143	&0.1954	&0.2269 \\
Specificity (True negative rate)			&0.9678	&0.9675	&0.9532 \\
F1 score									&0.3236	&0.2990	&0.3261 \\
AUC											&0.6837	&0.6804	&0.6767 \\
Matthews correlation coefficient			&0.2921	&0.2682	&0.2658 \\ \hline
\end{tabular}
}
\caption{Friendship detection with CFS feature selection on the temporal block assignment training data.}\label{tab:perf4}
\end{table}

While the test MCC of CV with unrestricted assignment goes down, with this fraction of only 19 features the test MCC of CV with dyadic assignment rises slightly, and the test MCC of CV with temporal block assignment does far better, going from an MCC of .21 to .27. These 19 features, then, seem to be picking up a significant portion of the pattern in co-location data, and a pattern that is more robust to changes over time. 

While it is dangerous to substantively interpret the selected features as causal or  even as necessarily stable \cite{mullainathan2017,yang2016}, it is a useful exploratory step to see the features that are effective for the detection task. The features are listed  in table (\ref{tab:selected}) ,with the pairwise correlations given in figure (\ref{fig:cormat}). While there are groups of highly linearly correlated features, many of the features are not correlated, giving an independent signal. 

There are some patterns that emerge in this well-performing subset of features. Threshold 2 (422m) shows up frequently, as do measures  related to variance (standard deviation  measures), nighttime, and the distribution of inverse squared distances. This generates several hypotheses: first, that  Latan\'{e} et al.'s \cite{latane1995} finding that inverse-squared distance fits well to reports of memorable social interactions may be effective for friendship detection as well. Second, the threshold at 422m seems particularly relevant versus others: this specific value might not be what is important, but perhaps this captures some relevant radius around the frat house. Otherwise, features associated with where people are co-located at night appear most frequently, which is in contrast to the finding by Eagle et al. \cite{eagle2009} that the daytime probability proximity is what was discriminative for friendships. 

\section{Conclusion and Future Work}
In this paper, we have described the collection of subjective, self-reported friendship data alongside objective sensor data within a given boundary specification. We modeled friendship, close friendships, and change in friendship with machine learning and evaluated them using three cross-validation schema that accounted for different use case scenarios in the real world to show the generalizability of our approach. We could detect friendship and close friendship with a significant better performance above baseline in both cases. Our change detection, however, performed poorly with current aggregated features, suggesting a different set  of features are needed to carry out this task. 

We also obtained a set of features through a CFS method on the most conservative training set (one constructed through temporal block assignment). Our test using the extracted features showed similar results to the full feature set, suggesting them as potential alternatives to the full feature set that can help building lightweight models, and suggesting that certain measures and timeframes, such as inverse squared distance, standard deviations, and nighttime patterns, are most helpful for detection. In our future work, we will further explore feature selection for a parsimonious set of features applicable for different detection tasks.  

Our findings demonstrate the feasibility of detecting friendships from location data, as well as establish the challenge of detecting {\it changes} in friendship. This opens possibilities for further investigating the relationship between friendship and co-location, as well as for designing mobile applications that build recommendation systems or interventions based on detected friendships. 

\bibliographystyle{ACM-Reference-Format}
\bibliography{malik_cohort_ubicomp2018}

\end{document}